\documentstyle[prb,aps,twocolumn,floats]{revtex}

\begin{document}
\input epsf
\draft
\wideabs{

\title {On the vortex motion in high temperature superconductors}

\author {I. L. Landau and H. R. Ott}
\address{Laboratorium f\"ur Festk\"orperphysik, ETH H\"onggerberg, 
CH-8093 Z\"urich, Switzerland}

\date{\today}
\maketitle

\begin{abstract}
We show that prominent features in voltage-current characteristics, recently measured 
in the mixed state of high-Tc superconductors and interpreted as evidence for an 
irreversibility line or a vortex-glass transition, may very well be explained with 
the simplest Kim-Anderson approach describing the vortex motion. In this case, the 
irreversibility line is not related to a transition in the system of vortices. 
Consulting numerous experimental reports on this subject we have not found a single 
example, which is in contradiction with this view. 
    
\end{abstract}
}    %wideabstract!!!

\section {Introduction}

The commonly accepted picture of the magnetic flux motion in high temperature 
superconductors (HTSC) is extremely complex. Many different theoretical models have 
been developed in order to explain various aspects of the flux-creep process (see 
Ref. 1 for details). The complexity of the models that have been invented to capture 
the motion of vortices in HTSC is related to the apparent inability of the simple 
Kim-Anderson approach for explaining the experimental results. \cite{2,3} However, a 
serious reconsideration of available data indicates that this may not really be the 
case. Recent experiments have shown that the low-temperature flux-creep data may 
perfectly well be described using the Kim-Anderson approach, if a realistic profile 
of the pinning potential is taken into account. \cite{4,5,6} In this paper we use a 
similar approach to analyze the motion of vortices at temperatures close to the 
superconducting critical temperature $T_{c}$. We show that the very specific features 
in voltage-current characteristics that manifest the flux-creep process and which are 
usually related to the irreversibility line and a vortex-glass transition, not only 
may be explained by employing the Kim-Anderson approach, but are a direct consequence 
of this simple model. First, we briefly recapitulate the model and its consequences 
and subsequently address some issues concerning the irreversibility 
line and the vortex-glass transition.

\section {The model}

If the current density $j$ in a sample is less than its critical value $j_{c}$, all 
vortices are pinned and their motion is only due to a thermally activated hopping of 
the vortex lines over potential barriers or via quantum tunneling through the 
barriers. The probability of hopping in the case of thermal activation is
%%%%
\begin{equation}
w=\nu _0\exp \left( {-{U \over {k_BT}}} \right),
\end{equation}
%%%%
where $\nu _0$ is the attempt frequency of the vortex line to surmount the potential 
barrier of height $U$ and $k_B$ is the Boltzmann constant. 
%%%%%%%%%%%%%%%%
\begin{figure}[!t]
 \begin{center}
  \epsfxsize=0.9 \columnwidth \epsfbox {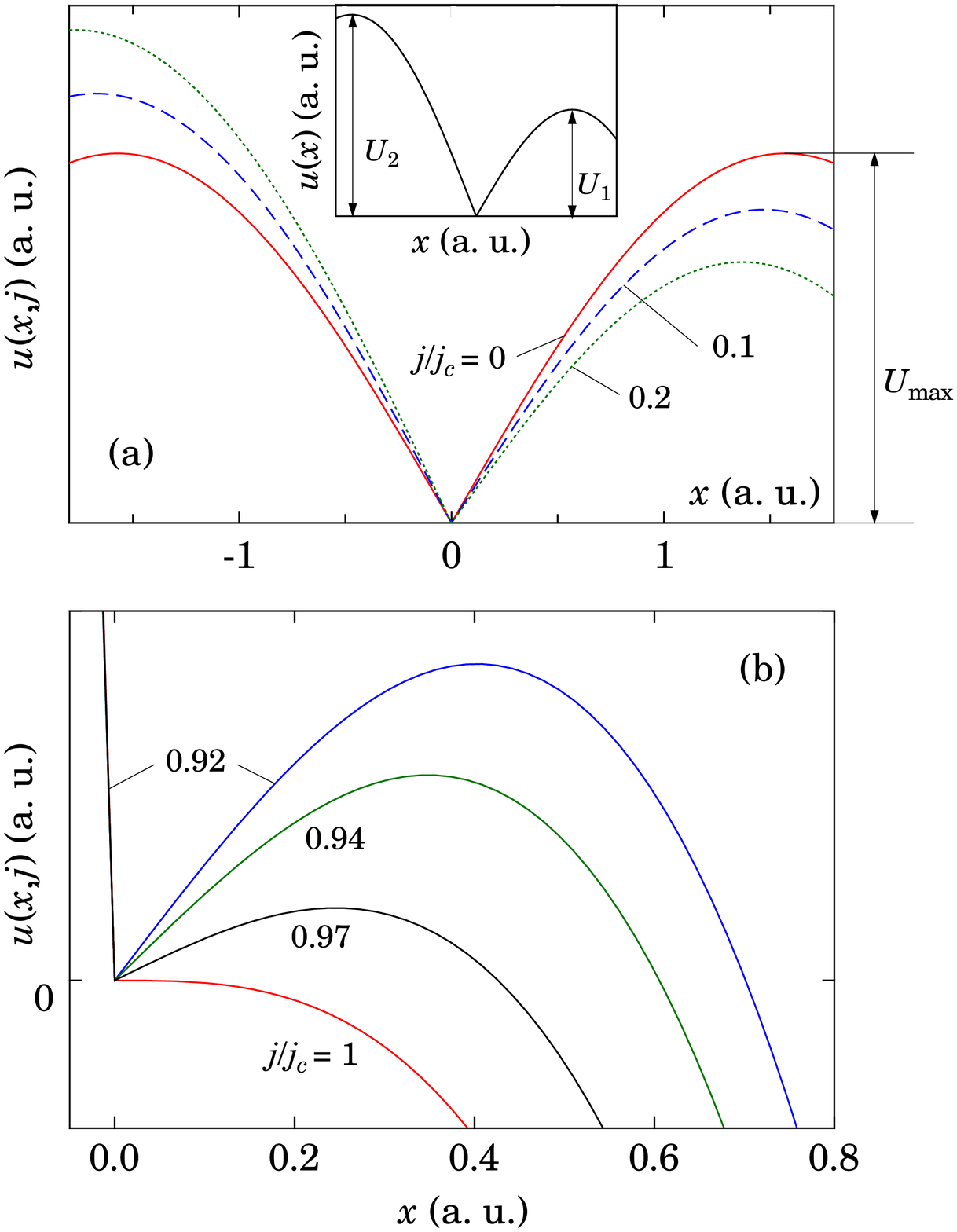}
  \caption{Schematic profiles of a pinning well for different values of the current 
           density. The corresponding values of $j/j_{c}$ are indicated near the 
		   curves. (a) $j/j_{c} \ll 1$. Inset illustrates definitions of $U_{1}$ and 
		   $U_{2}$. (b) $1 - j/j_{c} \ll 1$.}
  \end{center}
\end{figure} 
%%%%%%%%%%%%%%%%

As an example, a pinning-potential profile proportional to $\left| {\sin x} \right|$ 
is shown by the solid line in Fig. 1(a). At a given temperature $T$, the profile of 
the potential well for $j = 0$ may be written as 
%%%%
\begin{equation}
u(x) = U(T)f(x,T),
\end{equation}
%%%%
where $U(T)$ is the energy fixing the pinning strength and the function $f(x,T)$ 
defines the shape of the potential well, which may be temperature dependent  as 
well. We have chosen $f$ such that $\left| {df/dx} \right|$ is maximum at $x = 0$, 
the site of the minimum of a single well. $U(T)$ decreases with increasing 
temperature and it vanishes at the temperature at which the external magnetic field 
$H$ is equal to the upper critical field $H_{c2}$. In the following, $U_{\max}$ 
denotes the maximum value of $u(x,j=0)$ between two local minima as is illustrated 
in Fig. 1(a). Also $U_{\max}$ is decreasing with increasing $T$ but its temperature 
dependence may differ from that of $U(T)$.

An electric current does not change the interaction of a vortex with the pinning 
centers or other vortices, but it causes a Lorentz force $F_{L}$ to act on the 
vortices. The Lorentz force tilts the potential profile, thus reducing the potential 
barriers in the direction of the vortex motion. In the presence of a current the 
potential profile may be written as
%%%%
\begin{equation}
	u(x,j)=u(x,0)-xF_L
\end{equation}
%%%%
with $F_{L} = j \Phi _{0}/c$, $j$ as the current density, $\Phi _{0}$ the magnetic 
flux quantum, and $c$ the speed of light. The critical current density is reached 
if the potential barriers vanish. According to Eqs. (2) and (3), this results in
%%%%
\begin{equation}
j_c(T)={{cU(T)f'_c} \over {\Phi _0}},
\end{equation}
%%%%
where $f'_c$ is the maximum value of $\left| {df/dx} \right|$. As stated above, 
$\left| {df/dx} \right|$ reaches its maximum at $x = 0$, i.e., $f'_c= \left| {df/dx} 
\right|_{x=0}$. Equation (4) is a formal definition of the critical current density, 
the only appropriate definition of $j_{c}$ that may be given in the mixed state of 
type-II superconductors. As will be shown in section IV (see Fig. 6a), at 
temperatures close to the superconducting critical temperature $T_{c}$, the sample 
resistance in the mixed state may be rather high even if $j \ll j_{c}$. This is the 
reason why different voltage criteria for the evaluation of the critical current 
density, although useful 
for practical purposes, are rather meaningless from the point of view of physics. 

The influence of current on the potential profile is illustrated in Figs. 1(a) and 
1(b). It may be seen that at low currents ($j \ll j_{c}$) the decrease of the 
activation energy with increasing current is entirely determined by the behavior of 
$u(x)$ near its maximum [see Fig. 1(a)], while at currents close to $j_{c}$ only 
$u(x)$ in the vicinity of $x = 0$ is important [see Fig. 1(b)]. We denote the 
heights of potential barriers in the direction of the Lorentz force and opposite to 
it by $U_{1}$ and $U_{2}$, respectively [see inset to Fig. 1(a)]. For a non-zero 
current, $U_{1} < U_{2}$. The electrical field $E$ in the sample is proportional to 
the average velocity of vortices and may be written as
%%%%
\begin{equation}
E=E_0\left[ {\exp \left( {-{{U_1(T,j)} \over {k_BT}}} \right)-\exp \left( 
{-{{U_2(T,j)} \over {k_BT}}} \right)} \right]
\end{equation}
%%%%
with
%%%%
\begin{equation}
E_0={B \over c}l\nu _0.
\end{equation}
%%%%
Here $B$ is the magnetic induction in the sample and $l$ is the vortex hopping 
distance. The second term in Eq. (5) describes the vortex hopping in the direction 
opposite to the Lorentz force. This term is only important for $j/j_{c} \ll 1$ and 
at high temperatures. \cite{7}

If the function $f$ in Eq. (2) is known, $U_{1}$ and $U_{2}$ may be deduced by using 
Eq. (3). In practice however, the situation is quite different. The main goal is 
rather to obtain information about $u(x)$ from experimental results. In most cases, 
the experimentally accessible information is limited to $E(j)$ curves, measured at 
different temperatures. In order to obtain any useful information about $u(x)$ from 
these experimental data, some \textit{a priori} assumptions have to be made. As has 
been shown recently, the flux-creep rates in an epitaxial YBa$_{2}$Cu$_{3}$O$_{7-x}$ 
(YBCO) film may be very well described in a  wide range of temperatures ($0.02 
T_{c} \le T \le 0.9 T_{c}$) with the assumption that not the shape, but only the 
amplitude of the $u(x)$ function is temperature dependent. \cite{4,5,6} In this 
case, the function $f$ in Eq. (2) depends only on $x$ and the temperature 
dependencies of $U_{\max}$ and $U(T)$ coincide. Based on this assumption, a scaling 
procedure for the analysis of flux-creep data has been established. \cite{4,5,6} It 
is an important advantage of this approach that it allows for internal consistency 
checks and all assumptions that have been made may be verified retrospectively. It 
turns out that the scaling procedure breaks down at $T \ge 80 K \approx 0.9 T_{c}$. 
\cite{6} This demonstrates that the assumption of a temperature independent $f$ 
function is not valid close to the critical temperature and therefore should be 
abandoned in this temperature range. 

In the following we use Eq. (5) to analyze the voltage-current characteristics of 
the sample at temperatures close to $T_{c}$. It will be shown that different 
consequences of the vortex motion, such as the vanishing of the persistent current 
at $T_{irr} < T_{c}$ (irreversibility line) and the often observed sign change in 
the curvature of the $\log E- \log j$ curves with increasing temperature, which is 
usually attributed to a vortex-glass transition, follow straightforwardly from Eq. 
(5) without any additional assumption. 

\section {Irreversibility line}

Among numerous unusual features of HTSC, the so-called irreversibility line (IRL) 
has attracted a lot of attention. This line in the $H-T$ phase diagram separates 
two regions with distinctly different behavior. \cite{8,9,10,11,12,13,14,15} Above 
the IRL the magnetization of the sample is reversible, which means that the sample 
cannot carry any persistent current. Below the IRL, irreversible magnetization is 
observed. Since the true superconducting state with zero dissipation is achieved 
only below the IRL, the position of this line in the $H-T$ diagram is extremely 
important for applications. However, in spite of many years of intensive studies, 
the situation with regard to the IRL is far from being clear. Not only the origin 
of the IRL is still under discussion, but there is also no consensus concerning 
valid procedures to establish the position of the IRL from the experimental data. 
%%%%%%%%%%%%%%%%
\begin{figure}[h]
 \begin{center}
  \epsfxsize=0.9 \columnwidth \epsfbox {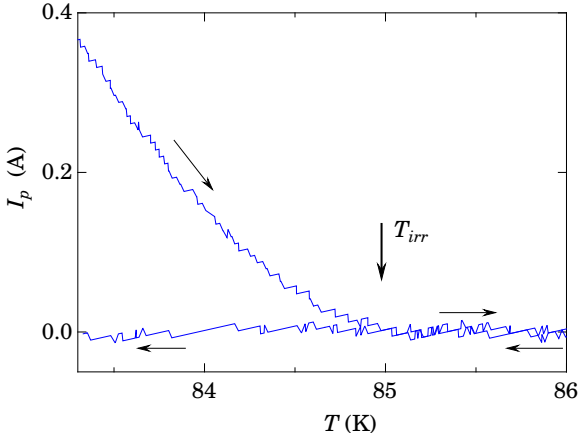}
  \caption{Persistent current $I_{p}$ in a ring-shaped YBa$_{2}$Cu$_{3}$O$_{7-x}$ 
           film as a function of temperature (description of the sample may be 
		   found in Ref. 4). The vertical arrow indicates the position of the 
		   irreversibility temperature as given by the present experiment. The 
		   sample was cooled in an external magnetic field $H \approx 1000$ Oe 
		   to $T = 82$ K. The current was subsequently induced by enhancing the 
		   magnetic field by 1 Oe.}
  \end{center}
\end{figure} 
%%%%%%%%%%%%%%%%

In order to make the physics more transparent, we consider here a ring-shaped 
sample. In this case the persistent current in the sample $I_{p}$ plays the role 
of an irreversible magnetic moment $M_{irr}$. The current in the ring may easily 
be monitored by measuring the magnetic induction in the ring cavity using, for 
instance, a Hall probe placed in the center of the cavity. Typical experimental 
data for a ring-shaped sample obtained from an epitaxial YBCO film are presented 
in Fig. 2. This figure shows a heating-cooling cycle of $I_{p}$. Our data in Fig. 
2, as well as results of measurements of the irreversible magnetization, 
\cite{8,9,10,11,12,13,14,15} reveal that above the irreversibility temperature, 
$T_{irr}$, the persistent current is essentially zero. This is why $T_{irr}$ is 
usually considered as the temperature, at which the critical current density 
vanishes. It is commonly accepted that $T_{irr}$ corresponds to some transition 
invoking the vortex system of the sample. Among a few considered possibilities 
for such a transition, the melting of a vortex-glass phase is the most popular 
explanation for the irreversibility line in HTSC. \cite{18} However, as shown 
below, this kind of temperature dependence of the persistent current $I_{p}$ 
necessarily follows from the simplest Kim-Anderson approach for describing the 
flux-creep process.

At temperatures well below the IRL the situation is quite clear. All vortices are 
pinned and their motion only occurs due to a thermally activated hopping of vortex 
lines over potential barriers. According to Eq. (1) the probability of hopping 
depends exponentially on $U/k_{B}T$. Equation (1) implies that the resistance of 
a type-II superconductor in the mixed state formally never vanishes. However, at 
low temperatures and small currents, the ratio $U/k_{B}T$ is so large that the 
probability of hopping is negligible. In this case, an electric current induced 
in a ring-shaped sample may flow without a noticeable decay for years. Because 
the ratio $U/k_{B}T$ decreases rapidly when approaching $T_{c}$, the current-decay 
rate increases significantly with increasing temperature. 

The current decay rate $dj/dt$ is proportional to the electrical field in the 
sample and therefore Eq. (5) may be used for evaluating this important quantity. 
In order to calculate the temperature and current dependencies of $dj/dt$, we have 
to assume some explicit expression for the profile of the potential well $u(x)$. 
Unfortunately, practically nothing is known about $u(x )$ in this high temperature 
range. In our previous study it was shown that the function $f$ in Eq. (2) may be 
considered as temperature independent only for $T < 0.9 T_{c}$. \cite{6} For 
higher temperatures the temperature dependence of $f$ has to be taken into account. 
In the following analysis we have chosen two rather different representations for 
$f(x,T)$, i.e.,
%%%%
\begin{equation}
f(x,T)=\left| x \right|-a(1-T/T_c)^kx^2
\end{equation}
%%%%
and
%%%%
\begin{equation}
f(x,T)=\left( {\sqrt {\left| x \right|+x_0}-\sqrt {x_0}} \right)-b{{\left( {\left| 
x \right|+x_0} \right)^{3/2}-x_0^{3/2}} \over {\left( {1-T/T_c} \right)^m}},
\end{equation}
%%%%
with
%%%%
\begin{equation}
U(T)=U_0(1-T/T_c)^{3/2}
\end{equation}
%%%%
for both cases. This kind of $U(T)$ follows from the simplest consideration of 
the vortex pinning at temperatures close to $T_{c}$. \cite{16,17} The potential 
profiles represented by Eqs. (7) and (8) are depicted in the insets to Fig. 3. 
We do not pretend that any of the $f$ functions given by Eq. (7) or (8) exactly 
represent the real situation in HTSC. Our choice of $f$ has been made in view of 
obtaining a reasonable profile shape and the possibility to perform analytical 
calculations. As will be shown below, the particular choice of $f$ does not 
influence the main qualitative features of the flux-creep process. 
%%%%%%%%%%%%%%%%
\begin{figure}[ht]
 \begin{center}
  \epsfxsize=0.9 \columnwidth \epsfbox {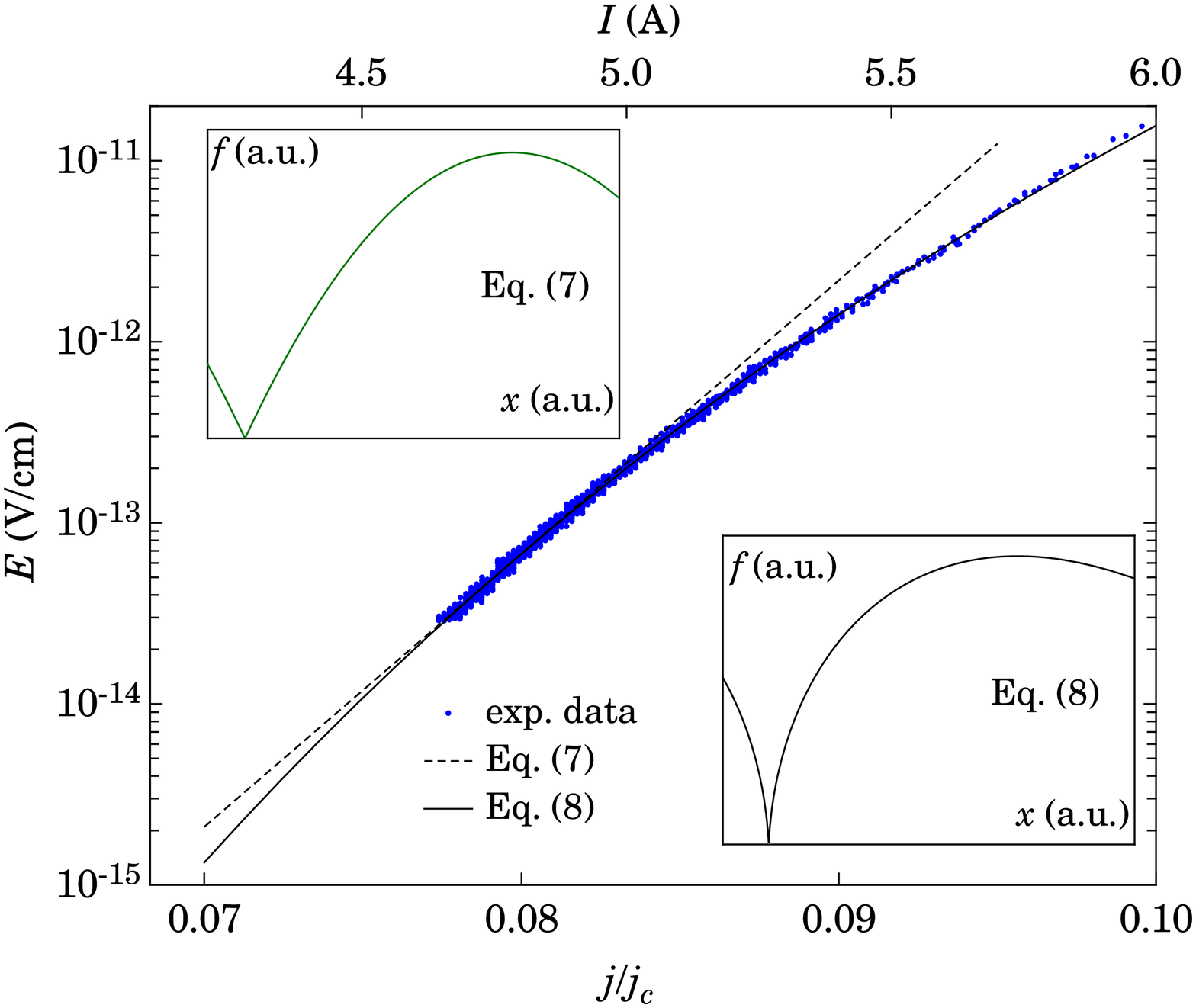}
  \caption{Experimental $E-j$ curve from Ref. 6 for $T = 0.9T_{c}$ and $H = 0.9$ 
           kOe. The lines are approximations to this curve using Eqs. (7) and (8) 
		   for the profile of the potential well. The upper scale shows absolute 
		   values of thr current. The upper and lower insets show examples of 
		   the profiles of the potential well as given by Eqs. (7) and (8), 
		   respectively.}
  \end{center}
\end{figure} 
%%%%%%%%%%%%%%%%
%%%%%%%%%%%%%%%%
\begin{figure}[!b]
 \begin{center}
  \epsfxsize=0.9 \columnwidth \epsfbox {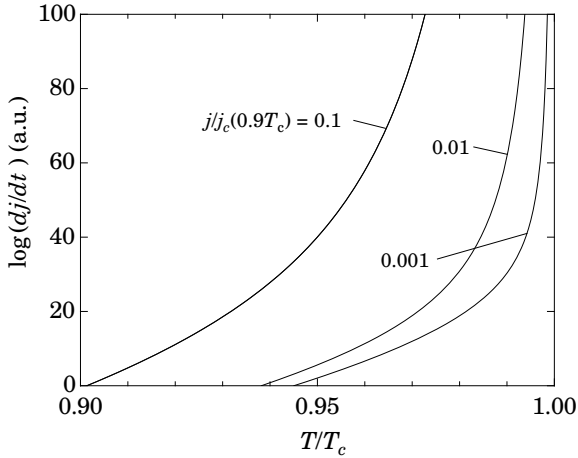}
  \caption{Calculated variation of $dj/dt$ versus $T/T_{c}$ for $f(x,T)$ given 
           by Eq. (7) with $k = 1.4$. Calculations were made for 3 fixed current 
		   densities. The values of the renormalized current densities $j/j_{c}
		   (0.9T_{c})$ are indicated near the curves.}
  \end{center}
\end{figure} 
%%%%%%%%%%%%%%%%
%%%%%%%%%%%%%%%%
\begin{figure}[ht]
 \begin{center}
  \epsfxsize=0.9 \columnwidth \epsfbox {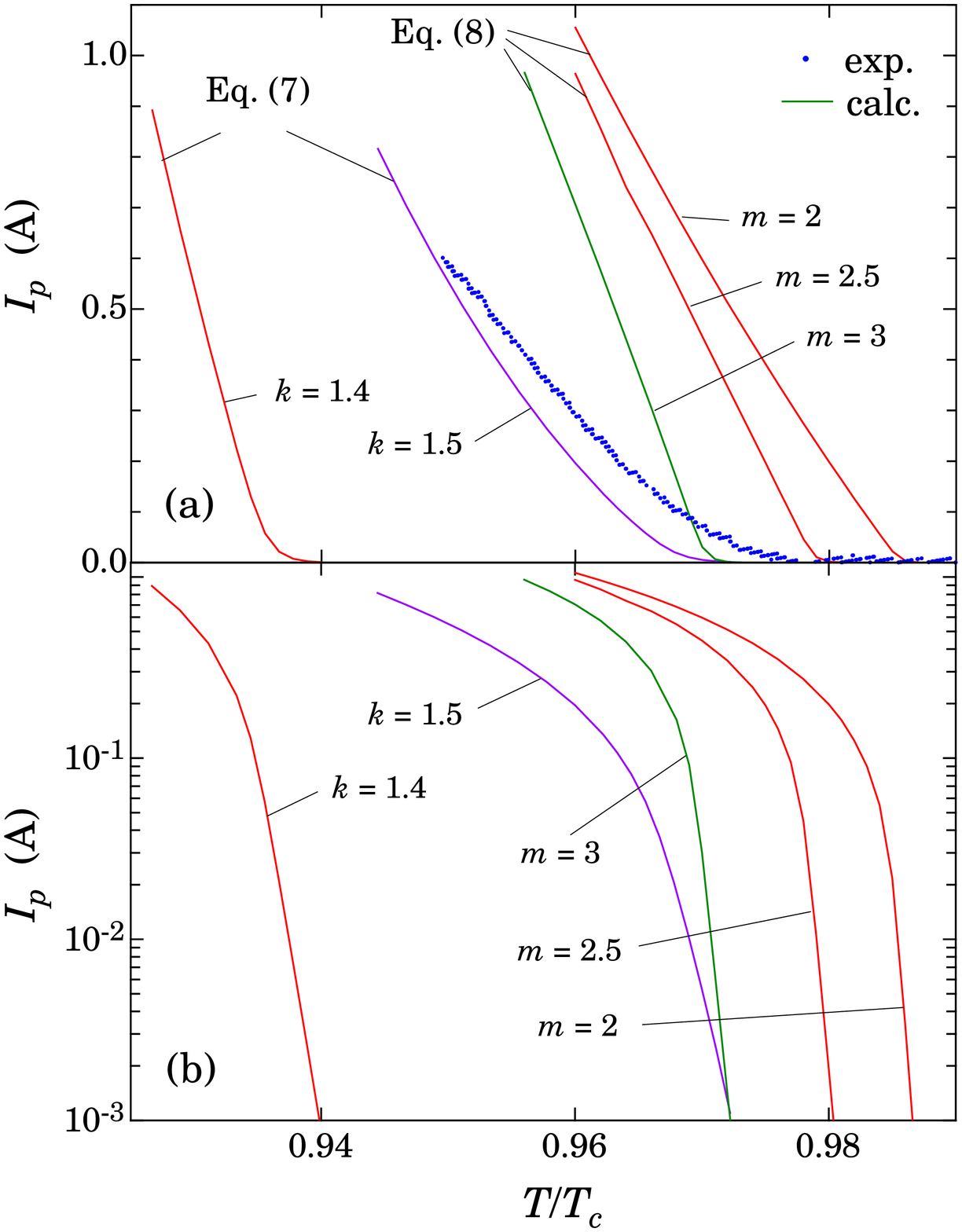}
  \caption{$I_{p}$ as a function of temperature. The solid lines are results of 
           calculations for different representations of $u(x)$. (a) On linear 
		   scales, the points represent the experimental data from Fig. 2. (b) 
		   $\log I_{p}$ versus $T/T_{c}$.}
  \end{center}
\end{figure} 
%%%%%%%%%%%%%%%%

We may now insert the potential profiles given by Eqs. (7) to (9) into Eq. (2) 
and calculate $U_{1}(T,j)$ and $U_{2}(T,j)$ with the help of Eq. (3). Substituting 
the resulting expressions for $U_{1}$ and $U_{2}$ into Eq. (5), we obtain the 
corresponding temperature and current dependencies of the electrical field $E$ in 
the sample. In order to compare the results of this calculation with the 
experimental data presented in Fig. 2, we need an evaluation of $E_{0}$ and 
$U_{0}$ entering Eqs. (5) and (9), respectively. For this purpose, we calibrated 
our calculated $E(j)$ via a voltage-current characteristic measured for the same 
sample at $T=T_{0} \approx 0.9 T_{c}$. \cite{6} Fig. 3 shows the experimental data 
together with the calculated $E(j)$ curves using $f(x,T)$ as given by Eqs. (7) and 
(8), respectively. In the first case, the constants $E_{0}$ in and $U_{0}$ were 
adjusted such as to approximate the low voltage part of the $E(j)$ curve; the 
constant $a=(1-T_{0}/T_{c})^{-k}$. In the second case, the parameters $x_{0}$ and 
$b$ in Eq. (8) were used as fitting parameters, as well. 

Using these adjusted expressions for $E$, we may now calculate different 
characteristics of the sample for the chosen $f(x,T)$. Fig. 4 shows the 
temperature dependencies of $dj/dt \propto E$ for three values of the current 
density. In this particular case, the calculation was made for $f(x,T)$ given 
by Eq. (7) with $k = 1.4$. Note that we use a log-scale for the $dj/dt$-axis and 
that the total change in $dj/dt$ is 100 orders of magnitude. This figure clearly 
demonstrates that $dj/dt$ grows extremely fast with increasing temperature close 
to $T_{c}$. This implies that a current which is practically constant in time at 
a certain temperature, will be decaying rather quickly at a slightly enhanced 
temperature. 

Experimentally the persistent current $I_{p}$ is usually determined as the current 
which is not decaying during the time of the experiment. This definition of the 
persistent current is not universal because the value of $I_{p}$ is obviously 
dependent on the experimental resolution and the time window of the particular 
experiment. In our approach we can calculate the temperature dependence of the 
persistent current by choosing $dI/dt$ as set by the available resolution of the 
experiment. The following calculations were made for $dI/dt = 3\cdot 10^{-4}$ A/s, 
which is the resolution for the experimental data presented in Fig. 2. In order to 
demonstrate how the $I_{p}(T) \propto M_{irr}(T)$ curve depends on the particular 
choice of the potential profile, the calculations were made for both 
approximations of $f(x,T)$ and for different values of the exponents $k$ and $m$ 
in Eqs. (7) and (8), respectively. The results of the calculations, together with 
experimental data from Fig. 2, are shown in Figs. 5(a) and 5(b). It may be seen 
that, depending on the degree of experimental resolution, an "irreversibility" 
temperature exists for all the chosen functions $f(x,T)$. At the same time, both 
the shape of the $I_{p}(T)$ curves and the position of $T_{irr}$ are rather 
sensitive to the choice of $f$. This means that by a proper choice of $U(T)$ and 
$f(x,T)$ in Eq. (2), any experimentally observed temperature dependence of the 
irreversible magnetic moment $M_{irr}(T)$ or the persistent current $I_{p}(T)$ 
may be approximated to a high degree of accuracy and no specific transition in 
the vortex system is needed to explain the existence of the irreversibility line 
established in this way. 

\section {Vortex-glass transition}

As has been mentioned above, the most popular interpretation of the 
irreversibility line in the $H-T$ diagram is that it is a manifestation of a 
vortex-glass transition. \cite{8,18,19} Many experimental results seem to confirm 
the concept of a vortex-glass melting at temperatures close to the IRL. 
\cite{12,20,21,22,23,24,25,26,27,28,29} In this section we show that the distinct 
variation of the shape of the $\log E- \log j$ curves, which is usually 
interpreted as a manifestation of the vortex-glass melting, follows 
straightforwardly from our simple consideration outlined in the previous section. 
%%%%%%%%%%%%%%%%
\begin{figure}[ht]
 \begin{center}
  \epsfxsize=0.9 \columnwidth \epsfbox {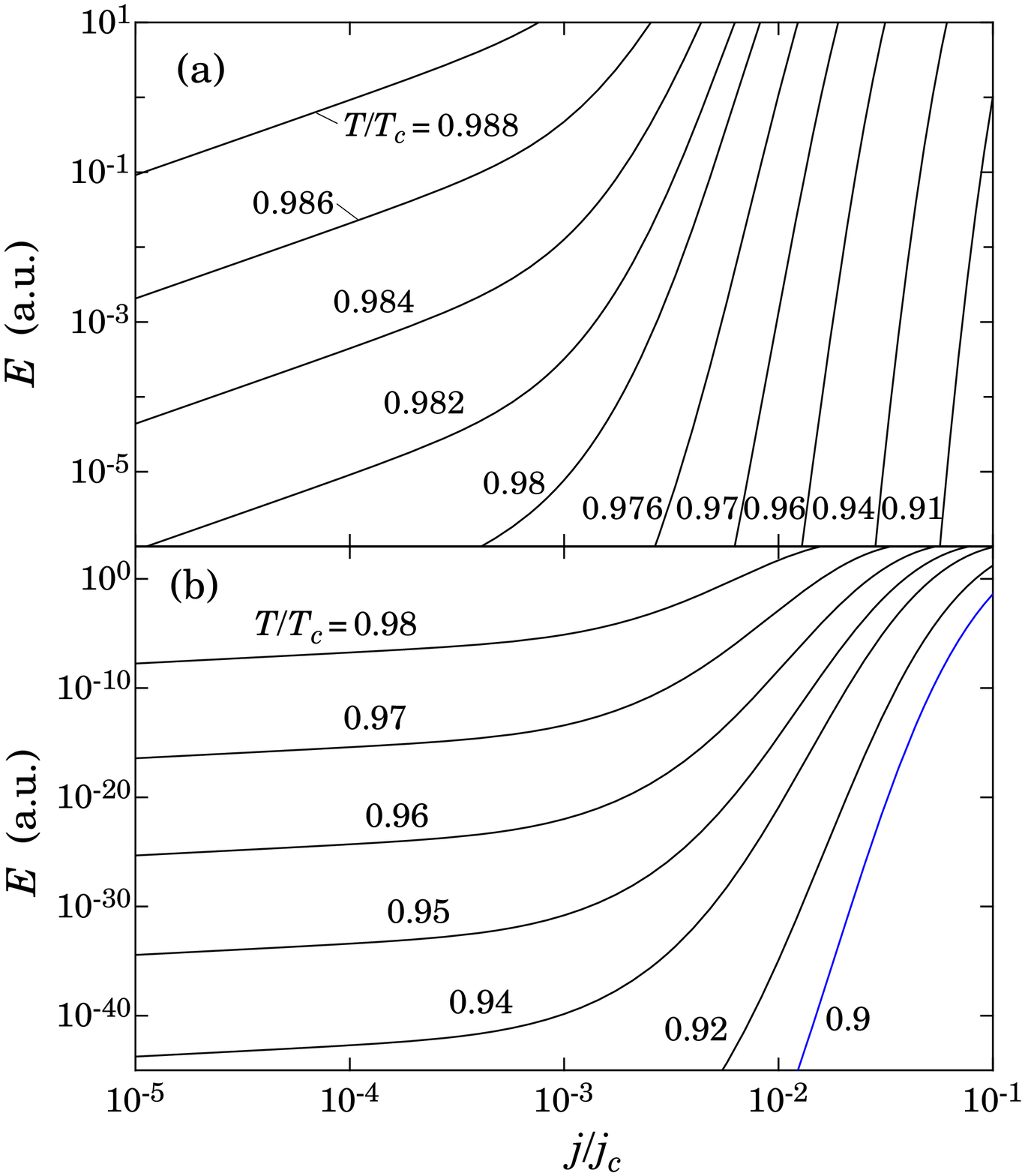}
  \caption{(a) and (b) $E-j$ curves calculated for $f(x,T)$ given by Eq. (8) with 
           $m = 2$ and parameters $x_{0}$ and $b$ detirmined as is explained in 
		   the previous section.}
  \end{center}
\end{figure} 
%%%%%%%%%%%%%%%%

To make the point, we use $f(x,T)$ as given by Eq. (8) with $m = 2$. The results 
of the calculations of $E(j)$ are shown in Fig. 6(a) as $E(j)$ curves at fixed 
temperatures. Qualitatively this plot is indistinguishable from numerous 
experimental results (see Refs. 20-29) and it is clear that by corresponding 
adjustments of $U(T)$ and $f(x,T)$, any experimental $E(j,T)$ curve may be 
approximated even quantitatively by this type of calculation. Fig. 6(b) shows the 
same $E(j)$ curves as Fig. 6(a) at much lower voltages. It may be seen that the 
change of the curvature, which is usually attributed to the vortex-glass 
transition, is a universal feature of the $E(j)$ curves at  temperatures close to 
$T_{c}$. With decreasing temperature however, the sign change of the curvature is 
shifted to lower voltages where it is not accessible experimentally. 

We emphasize that the crossover to ohmic $E(j)$ curves with decreasing current 
is independent of the particular choice of the $u(x)$ function. As has been 
demonstrated in Section II, for $j/j_{c} \ll 1$, only a small part of the $u(x)$ 
function near its maximum is important. In this region, $u(x,0)$ may be replaced 
by a Taylor expansion of the form 
%%%%
\begin{equation}
u(x)=U_{\max }\left[ {1-{{A^2} \over 2}\left( {{{\left| x \right|} \over {x_0}}-1} 
\right)^2} \right].
\end{equation}
%%%%
Here $x_{0}$ is the point where $u(x,0)$ has its maximum and $A$ is the curvature 
of $u$ at $x = x_{0}$. Both $A$ and $x_{0}$ may be temperature dependent. Using 
Eqs. (3) and (10), the activation energies are 
%%%%
\begin{equation}
U_{1,2}=U_{\max }\left[ {1\mp j{{x_0\Phi _0} \over {cU_{\max }}}+{{j^2} \over 2} 
\left( {{{x_0\Phi _0} \over {cAU_{\max }}}} \right)^2} \right].
\end{equation}
%%%%
Inserting (11) into Eq. (5), we get
%%%%
\begin{equation}
E=E_0\exp \left( {-{{U_{\max }} \over {k_BT}}} \right)\exp \left[ {{{(jx_0\Phi 
_0/cA)^2} \over {2k_BTU_{\max }}}} \right]\sinh \left( {{{jx_0\Phi _0} \over 
{ck_BT}}} \right).
\end{equation}
%%%%
Because of the additional exponential term, the $j$ dependence of $E$ cannot be 
reduced to a hyperbolic sinus. \cite{7} Taking into account that Eq. (12) is only 
valid at the low-current limit, we may use the expansions of the exponential and 
hyperbolic sinus functions and neglect all terms $j^{n}$ ($n \ge 2$). Eq. (12) is 
thus reduced to 
%%%%
\begin{equation}
E=j{{x_0\Phi _0} \over {ck_BT}}E_0\exp \left( {-{{U_{\max }} \over {k_BT}}} 
\right)
\end{equation}
%%%%
and hence at any temperature, the sample resistivity $\rho = E/j$ is independent 
of the current 
density for $j \ll j_{0} = cU_{\max}/x_{0} \Phi_{0}$. 

A very similar explanation of the voltage-current characteristics near the 
"vortex-glass" transition has been suggested by Coppersmith at al.. \cite{30} In 
their short comment they considered a sinusoidal potential barrier. It was shown 
that even with this simple potential all the qualitative features of the 
experimental $E(j)$ curves could be reproduced. The authors also pointed out that 
the insignificant quantitative disagreement with the experimental data is simply 
due to the arbitrary chosen sinusoidal profile of the potential barriers. 

\section {Conclusion}

As has been shown in Sections III and IV, the experimentally established 
vanishing of the persistent current at $T = T_{irr} < T_{c}(H)$ does not 
necessarily mean that the critical current density vanishes at $T \ge T_{irr}$ 
and no singular event, such as a vortex-glass transition, needs to be involved 
in order to explain the experimentally verifiable disappearance of the 
irreversible magnetization at the irreversibility line. This feature, as well as 
the change of curvature of the voltage-current characteristics with temperature, 
which is usually considered as a manifestation of the vortex-glass melting, 
automatically follow from the simplest consideration of thermally activated 
vortex motion (see Figs. 5 and 6). 

In experimental papers dealing with vortex-glass melting, there is usually no 
discussion of alternative possibilities to account for the experimental results. 
All the data are scaled according to the vortex-glass model and the corresponding 
correlation lengths are determined. A common statement in most of these papers is 
that the flux-creep model cannot provide a negative curvature for the low 
temperature $\log E- \log j$ curves. \cite{22,23} As may be seen in Figs. 6(a) 
and 6(b), a negative curvature may very well be accounted for by the flux creep 
model. In a treatment of thermally activated vortex hopping given in Section 
IV, this curvature depends on the particular pinning potential.


\begin{references}
\bibitem{7} For any smooth potential profile u(x), Eq. (5) cannot be reduced to 
$E = E_{0} \exp (-U/k_{B}T) \sinh ([(U/k_{B}T)(j/j_{c})])$. The latter expression 
is based on a linear dependence of the activation energy on $j$ and is widely 
used in the literature. \cite{1,31} However, it is valid for triangular potential 
barriers only (see Refs. 4 and 6 for details).
\end{references}
\end{document}